\documentclass[pre,showpacs,twocolumn]{revtex4}
\usepackage{graphicx}
\begin{document}

\title{Validation of the Jarzynski relation for a system with strong thermal coupling: an isothermal ideal gas model}

\author{A. Baule, R. M. L. Evans, P. D. Olmsted}
\affiliation{School of Physics and Astronomy,
University of Leeds, 
LS2 9JT,
United Kingdom}

\begin{abstract}

We revisit the paradigm of an ideal gas under isothermal conditions. A moving piston performs work on an ideal gas in a container that is strongly coupled to a heat reservoir. The thermal coupling is modelled by stochastic scattering at the boundaries. In contrast to recent studies of an adiabatic ideal gas with a piston [R.C. Lua and A.Y. Grosberg, \textit{J. Phys. Chem. B} 109, 6805 (2005); I. Bena et al., \textit{Europhys. Lett.} 71, 879 (2005)], container and piston stay in contact with the heat bath during the work process. Under this condition the heat reservoir as well as the system depend on the work parameter $\lambda$ and microscopic reversibility is broken for a moving piston. Our model is thus not included in the class of systems for which the non-equilibrium work theorem has been derived rigorously either by Hamiltonian [C. Jarzynski, \textit{J. Stat. Mech.} P09005 (2004)] or stochastic methods [G.E. Crooks, \textit{J. Stat. Phys.} 90, 1481 (1998)]. Nevertheless the validity of the non-equilibrium work theorem is confirmed both numerically for a wide range of parameter values and analytically in the limit of a very fast moving piston, i.e. in the far non-equilibrium regime.

\end{abstract}


\maketitle

\def\d{{\rm d}}

\section{Introduction}

Despite its successful application in numerous experimental and numerical studies (see e.g. the review articles \cite{Ritort,Park}), the validity of the non-equilibrium work theorem or `Jarzynski relation' \cite{Jarzynski} still remains under discussion. Indeed this ongoing critique is mainly due to the surprising nature of the theorem: it states an exact equality that holds in situations arbitrarily far from equilibrium, under very general assumptions. More precisely, it states that the free energy difference between two equilibrium states can be extracted from work measurements along irreversible trajectories connecting these two states. Therefore one can, in principle, obtain equilibrium information from a non-equilibrium experiment, which is of particular interest in chemical and biophysical applications. For example, the non-equilibrium work theorem has been successfully applied to the stretching of a single protein \cite{Liphardt}. In this experiment the work performed by a single RNA molecule tethered between a solid substrate and a controllable cantilever in an aqueous salt solution is measured for slow (reversible) and fast (irreversible) stretching. The free energy difference between its folded and unfolded conformations is obtained from the reversible process using ordinary equilibrium thermodynamics. On the other hand applying Jarzynski's relation to the work values obtained from the irreversible process also reproduces this result within experimental errors, thus confirming the theorem.
 
However it has been questioned, in Ref.~\cite{Cohen}, whether this experiment indeed creates a non-equilibrium situation. It is argued that a slow or fast work process does not necessarily guarantee its reversibility or irreversibility. Rather the work rate has to be compared with the strength of the coupling (rate of heat transfer) between the system and its thermal environment. If the work rate is apparently large, but still smaller than the rate of heat transfer, the system is essentially maintained in an equilibrium state. This is claimed~\cite{Cohen} to be the case in the protein stretching experiment, since the surrounding liquid allows for rapid thermalization. Under such conditions the theorem is expected to hold trivially.
 
The above discussion highlights the importance of properly assessing the thermostating process between the system and the heat reservoir. The purpose of the present paper is to investigate the Jarzynski relation for the most simple thermodynamic system under isothermal and non-equilibrium conditions. In a gedankenexperiment an ideal gas is isothermally expanded in a heat-conducting container by pulling a piston at different velocities. Work is performed when the gas particles hit the piston during its movement. Similar ideal gas models have been investigated in \cite{Lua,Bena}, but under adiabatic conditions, i.e. without considering heat transfer during the work process. The extension to an isothermal situation provides important further insight into Jarzynski's relation.
 
The remainder of this paper is organized as follows. In the next section we present a brief outline of the non-equilibrium work theorem for a system with strong thermal coupling. In Sec.~\ref{ideal_gas} the isothermal ideal gas model is introduced, which allows for an analytical formulation of the non-equilibrium work theorem. We present the results of a numerical study of this model and revisit the adiabatic expansion of the ideal gas in Sec.~\ref{numerics}. Finally we conclude with a summary of the main points and a brief outlook.

\section{The non-equilibrium work theorem \label{work_theorem}}

The non-equilibrium work theorem can be formulated in the following way. The system of interest is prepared in an initial state of equilibrium while in contact with a heat reservoir at temperature $T$. By changing an external parameter $\lambda$ --- the work parameter --- according to a fixed protocol from A to B the system is subjected to a thermodynamic process at the end of which it reaches a final state not necessarily in equilibrium. This process can possibly drive the system arbitrarily far away from equilibrium while performing a certain amount of work, $W$. During the work process the system may or may not stay in contact with the thermal environment. Upon reaching the final parameter value $\lambda=B$ the system relaxes to equilibrium by exchanging heat with the reservoir but it is assumed that no further work is performed. If we repeat this process following the same protocol infinitely many times we obtain a distribution of work values $p(W)$ due to the stochastic nature of the initial equilibrium state, which is sampled from a canonical distribution. The Jarzynski relation then states a strong constraint on this work distribution \cite{Jarzynski}:
\begin{eqnarray}
\label{Jarzynski_eq}
\left<e^{-\beta W}\right>=\int \d W\;p(W)\;e^{-\beta W}=e^{-\beta \Delta F} .
\end{eqnarray}
The average over the exponentiated work values equals the exponential of the free energy difference between the initial and final equilibrium states of the system, whether or not the final equilibrium state is actually realized in the disturbed system. $\beta$ is the inverse temperature $1/\beta=k_BT$ and $\Delta F$ is the ratio of the equilibrium partition functions:
\begin{eqnarray}
\Delta F=F_B-F_A=-\beta^{-1}\ln\frac{Z_B}{Z_A} .
\end{eqnarray}
It should be noted that $F_B$ corresponds to the actual free energy of the final state only if the work parameter is finally held fixed at $\lambda=B$ until the system has thermalized with the reservoir. Without thermalization the system may well be in a state out of equilibrium such that no free energy can be defined.
 
The relation, Eq.~(\ref{Jarzynski_eq}), holds irrespective of the particular character of the work process and is valid beyond the linear response regime. For a fast switching of $\lambda$ one may perform a non-equilibrium process and still obtain the equilibrium free energy difference by evaluating the exponential work average. Thus the Jarzynski relation is one of the few exact results applicable far from equilibrium. Since $\left<\exp(x)\right>\geq \exp(\left< x\right>)$, Eq.~(\ref{Jarzynski_eq}) implies the second law of thermodynamics formulated for work and free energy, $\left<W\right>\geq\Delta F$. The equality $\left<W\right>=\Delta F$ is only true for a reversible (quasistatic) process.

The Jarzynski relation has been derived both for deterministic Hamiltonian dynamics \cite{Jarzynski,Jarzynski2} and for stochastic dynamics \cite{Crooks}. Without resorting to the full proof we shall make a few comments on the derivation of Eq.~(\ref{Jarzynski_eq}) for a system with strong thermal coupling.

\subsection{Hamiltonian derivation}

In Jarzynski's original derivation \cite{Jarzynski} almost a decade ago, the specific assumption was made that the coupling between system and heat reservoir is sufficiently small to neglect the interaction term in the Hamiltonian. Under this condition the work process is effectively assumed to be adiabatic. Only recently a derivation has been presented that does not rely on the weak coupling assumption \cite{Jarzynski2}. The starting point is the Hamiltonian
\begin{eqnarray}
\mathcal{H}(\Gamma;\lambda)=H(x;\lambda)+H_E(y)+h_{int}(x,y) ,
\end{eqnarray}
for the system and thermal environment. $H(x;\lambda)$ denotes the Hamiltonian of the system, $H_E(y)$ is the Hamiltonian of the heat reservoir and $h_{int}(x,y)$ is the interaction Hamiltonian. Here, $x$ refers to a point in the phase space of the system only, likewise $y$ for the heat reservoir. We denote a point in the combined phase space by $\Gamma=(x,y)$. It is important to note that the dependence on the work parameter enters only via the Hamiltonian of the system, while the heat reservoir is assumed to be $\lambda$-independent. The \textit{combined} system-plus-reservoir is now subject to an adiabatic work process $\lambda(t)$, such that the non-equilibrium work theorem can be applied in its original form. For finite $h_{int}(x,y)$ the equilibrium distribution of the system has to be described by the modified Boltzmann-factor $p_S(x;\lambda)\propto \exp[-\beta H^*(x;\lambda)]$ where $H^*(x;\lambda)$ is called a potential of mean force \cite{Jarzynski2,Park}:
\begin{eqnarray}
&&H^*(x;\lambda)\nonumber\\
&&=H(x;\lambda)-\beta^{-1}\ln\frac{\int \d y\exp\left[-\beta(H_E(y)+h_{int}(x,y))\right]}{\int \d y\exp[-\beta H_E(y)]} \nonumber\\
\end{eqnarray}
With this consideration, the left hand side of Eq.~(\ref{Jarzynski_eq}) is reduced to the ratio of the partition functions of the system only, resulting in the same result as in the case of weak coupling.
 
One should note a subtlety that applies to systems with rigid boundaries whose position varies with the control parameter $\lambda(t)$ (which is the case e.g.~for the free expansion of an ideal gas in a box, or expansion against a piston). Such boundaries have to be regarded as potentials in the Hamiltonian of the system. If one imposes instead time-dependent constraints, the resulting Hamiltonian evolution does not conserve phase space volume thus leading to an apparent violation of Jarzynski's relation. The correct procedure assumes a potential strength depending on a parameter $\epsilon$ which in the limit $\epsilon\rightarrow\infty$ becomes a rigid boundary. As a consequence evaluating the exponential work average in Eq.~(\ref{Jarzynski_eq}) requires correct ordering of the thermodynamic limit (number of repetitions) before the $\epsilon$ limit \cite{Silbey}.

\subsection{Derivation for stochastic microscopic reversible dynamics \label{CrooksDeriv}}

In \cite {Crooks} the non-equilibrium work theorem was derived under the assumption of stochastic Markovian microscopically reversible dynamics. The crucial condition of microscopic reversibility is formulated as \cite{Crooks}:
\begin{eqnarray}
\label{micro_rev}
\frac{P(x(t)|\lambda(t))}{P(\bar{x}(-t)|\bar{\lambda}(-t))}=\exp[-\beta Q(x(t),\lambda(t))]. 
\end{eqnarray} 
Here, $x(t)$ is a particular trajectory of the system when it is subject to the external work process $\lambda(t)$. $P(x(t)|\lambda(t))$ is the probability of following this path during the work process and $P(\bar{x}(-t)|\bar{\lambda}(-t))$ is the probability of the corresponding time-reversed path during the time-reverse of the process. $Q$ denotes the heat transferred from the heat reservoir to the system during this process; it is a functional of the path, with the property $Q(x(t),\lambda(t))=Q(\bar{x}(-t),\bar{\lambda}(-t))$ under time reversal. Equation~(\ref{micro_rev}) is claimed to hold arbitrarily far from equilibrium and allows for a simple proof of Jarzynski's relation as discussed in \cite{Crooks,Crooks2}. In the following we discuss the derivation of Eq.~(\ref{micro_rev}) more thoroughly.
 
Microscopic reversibility is usually expressed by the condition of detailed balance for the transitions between states $a$ and $b$ \cite{Kampen}:
\begin{eqnarray}
\label{db}
\frac{\omega(a\rightarrow b)}{\omega(b\rightarrow a)}=e^{-\beta(E_b-E_a)},
\end{eqnarray}
where the `states' $a$, $b$ refer to discrete volume elements in the phase space of the system and $\omega(a\rightarrow b)$, $\omega(b\rightarrow a)$ are the corresponding transition rates. It is argued in \cite{Crooks} that Eq.~(\ref{micro_rev}) follows directly by discretizing the path into single time steps, which each obey detailed balance. More precisely, the evolution of the system $x(t)=\{x_0,...,x_t\}$ is considered for a fixed sequence of the work parameter $\lambda(t)=\{\lambda_1,...,\lambda_t\}$ such that the single time steps can be decoupled into two substeps. First the control parameter is changed, $\lambda_i\rightarrow \lambda_{i+1}$, performing a certain amount of work, and then the system evolves for fixed $\lambda_{i+1}$, $x_i\rightarrow x_{i+1}$, exchanging heat with the thermal environment. As a consequence of Markovian dynamics the probability $P(x(t)|\lambda)$ of following a path through phase space under the work process can be expressed as a product of transition probabilities for the discretized quantities $\{x_0,...,x_t\}$ and $\{\lambda_1,...,\lambda_t\}$ \cite{Crooks}:

\begin{eqnarray}
P(x(t)|\lambda(t))=\prod_{i=0}^{t-1} P(x_i\rightarrow x_{i+1}|\lambda_{i+1}).
\end{eqnarray}
Then the ratio of probabilities of a forward path and its corresponding time-reversed path is
\begin{eqnarray}
\label{crooks_db}
\frac{P(x(t)|\lambda(t))}{P(\bar{x}(-t)|\bar{\lambda}(-t))}&=&\prod_{i=0}^{t-1} \frac{P(x_i\rightarrow x_{i+1}|\lambda_{i+1})}{P(x_i\leftarrow x_{i+1}|\lambda_{i+1})}\nonumber \\
&=&e^{-\beta\sum_{i=0}^{t-1}(E(x_{i+1},\lambda_{i+1})-E(x_{i},\lambda_{i+1}))}\nonumber\\
&=&e^{-\beta Q(x(t),\lambda(t))}.
\end{eqnarray}
In the second line it is assumed that, for each time step, a `detailed balance like' condition holds for the ratios of forward and reverse probabilities analogous to Eq.~(\ref{db}) {\em for fixed values of the work parameter} $\lambda$. The third line is due to the first law of thermodynamics: the difference in energy between two successive states is completely supplied by the heat bath if no work can be performed (i.e. for constant $\lambda$), $\Delta Q=E(x_{i+1},\lambda_{i+1})-E(x_{i},\lambda_{i+1})$. Similarly the work performed by the system originates only from a change in $\lambda$: $\Delta W=E(x_{i},\lambda_{i+1})-E(x_{i},\lambda_{i})$. This decoupling is crucial for the derivation of Eq.~(\ref{micro_rev}). By considering only transitions $x_i\rightarrow x_{i+1}$ for constant work parameter $\{\lambda_1,...,\lambda_t\}$ in Eq.~(\ref{crooks_db}), the evolution of the system is effectively reduced to a sequence of static states that obey detailed balance. Crucial to this is the assumption that the transition rates are independent of the rate $\dot{\lambda}$ at which the system is disturbed. One might therefore question whether Eq.~(\ref{micro_rev}) is indeed valid away from equilibrium, where detailed balance is not generally expected to hold \cite{Klein,Evans}. If detailed balance is violated under particular conditions, then Eq.~(\ref{micro_rev}) also fails. It is thus reasonable to state that the derivation of Eq.~(\ref{micro_rev}) is correct under the given assumptions, but that these assumptions do not properly take into account a non-equilibrium evolution of the system.
 
In the next section we will further discuss microscopic reversibility with regard to the isothermal ideal gas model.

\section{The ideal gas with a piston: an isothermal model \label{ideal_gas}}

\begin{figure}
\begin{center}
\includegraphics[width=6cm]{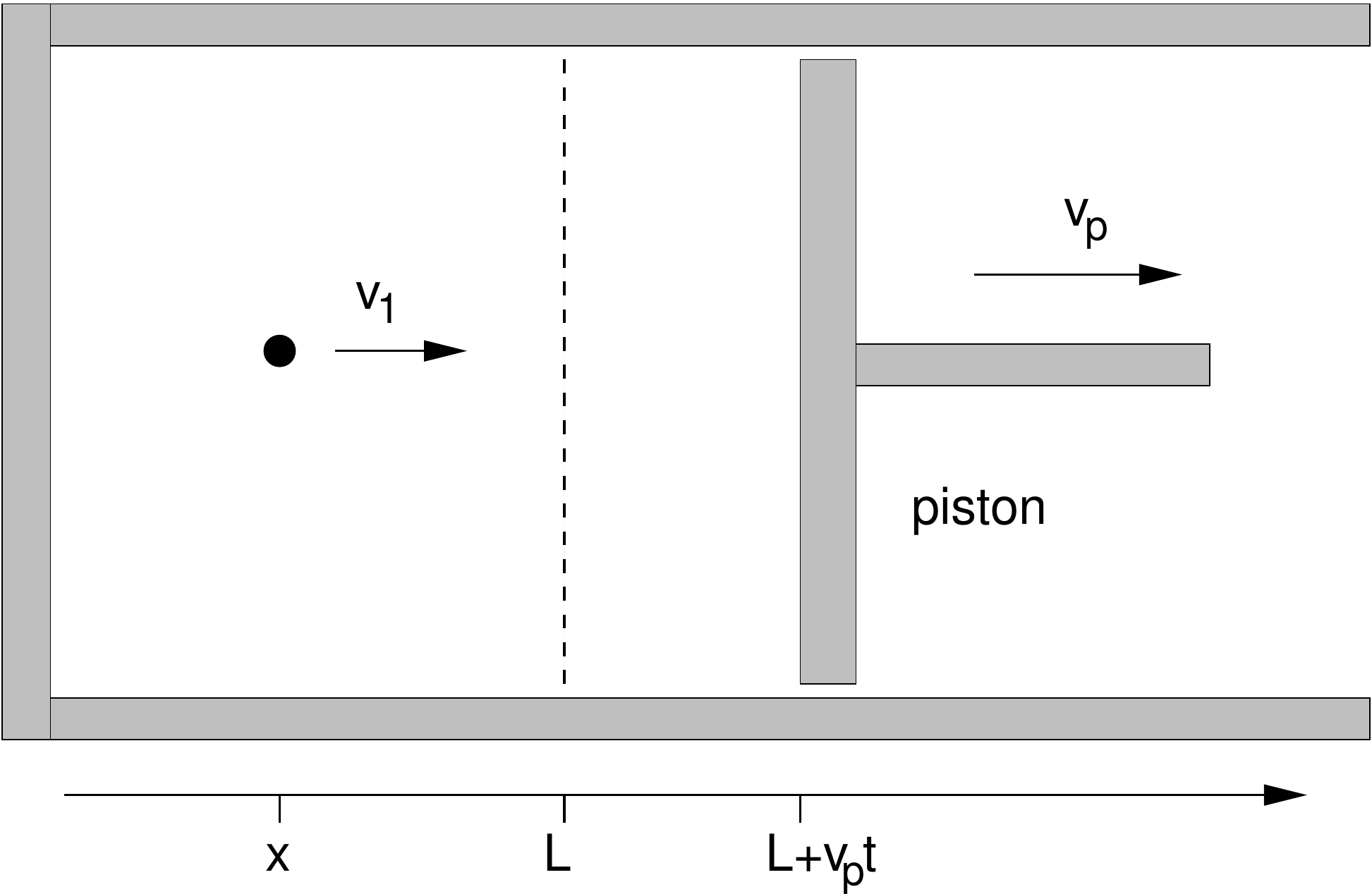}
\caption{\label{Piston}The ideal gas confined in a container with a piston. The initial position and velocity of a single gas particle are denoted $x$ and $v_1$, and the piston has velocity $v_p$. The (one-dimensional) gas is initially confined to a length $L$.}
\end{center}
\end{figure}

We consider a one-dimensional classical non-interacting ideal gas in a container with a moving piston (as shown in Fig.~\ref{Piston} which resembles the system treated in reference \cite{Lua}). Both the end wall of the container and the piston are connected to a heat reservoir which keeps the gas at constant temperature. This heat reservoir is assumed to be in thermal equilibrium such that its degrees of freedom are distributed according to a canonical Boltzmann distribution. Interactions between the system and its thermal environment are modeled by stochastic scattering at the boundaries: when the gas particle reaches either side of the container a completely inelastic collision takes place and the particle loses all its kinetic energy. It receives a new stochastic velocity which is sampled from the probability distribution of the heat reservoir, independent of its former velocity. We refer to this situation as a strong thermal coupling between system and environment.

Due to flux conservation, the probability distribution of the new particle velocity $\textbf{v}_{out}$ after a particular collision with the fixed (left-hand) boundary (proportional to the flux of particles leaving the boundary) therefore takes the form \cite{Lebowitz}:
\begin{eqnarray}
\label{rhoB}
\rho_{B}(v_{out})=\frac{|\textbf{v}_{out}|}{k_BT} \exp\left(-\frac{\textbf{v}_{out}^2}{2k_BT} \right).
\end{eqnarray}
Here $v_i$ always denotes the modulus $v_i\equiv|\textbf{v}_i|$. Under equilibrium conditions, this boundary condition yields the Boltzmann distribution for the velocities of particles in the container volume. 
A similar expression applies to the distribution of velocities assigned at the moving boundary (the piston), but with the important distinction that this is the distribution of out-going velocities {\em in the frame of the piston}. Movement of the piston therefore results in a non-zero mean streaming velocity in the laboratory frame.

In contrast to the commonly used Gaussian or Nos\'e-Hoover thermostating schemes, this stochastic boundary thermostat is non-deterministic and non-time-reversible (in the lab frame). Nevertheless it provides a valid physical model of the heat bath interaction. Furthermore, since no potential acts on the ideal gas particles, their energy is purely kinetic and completely determined by the canonical probability distribution of the heat reservoir.

An important property of the heat reservoir in this model is its dependence on the work parameter $\lambda$, which is more precisely a $\lambda$-dependence of its center of mass. In the counter-intuitive context of the Jarzynski equation this should not be considered trivial. As has been mentioned in the previous section, the derivation of the non-equilibrium work theorem assumes $\lambda$-dependence only for the Hamiltonian of the system, not for the heat reservoir. Here, that assumption is violated by the moving piston, which, by definition, changes its position as a function of $\lambda$ and, at the same time, thermostats the system. Although a rigorous and general treatment of this issue would require a Hamiltonian description of the heat reservoir, the stochastic model provides important insight.

\begin{figure}
\begin{center}
\includegraphics[width=6cm]{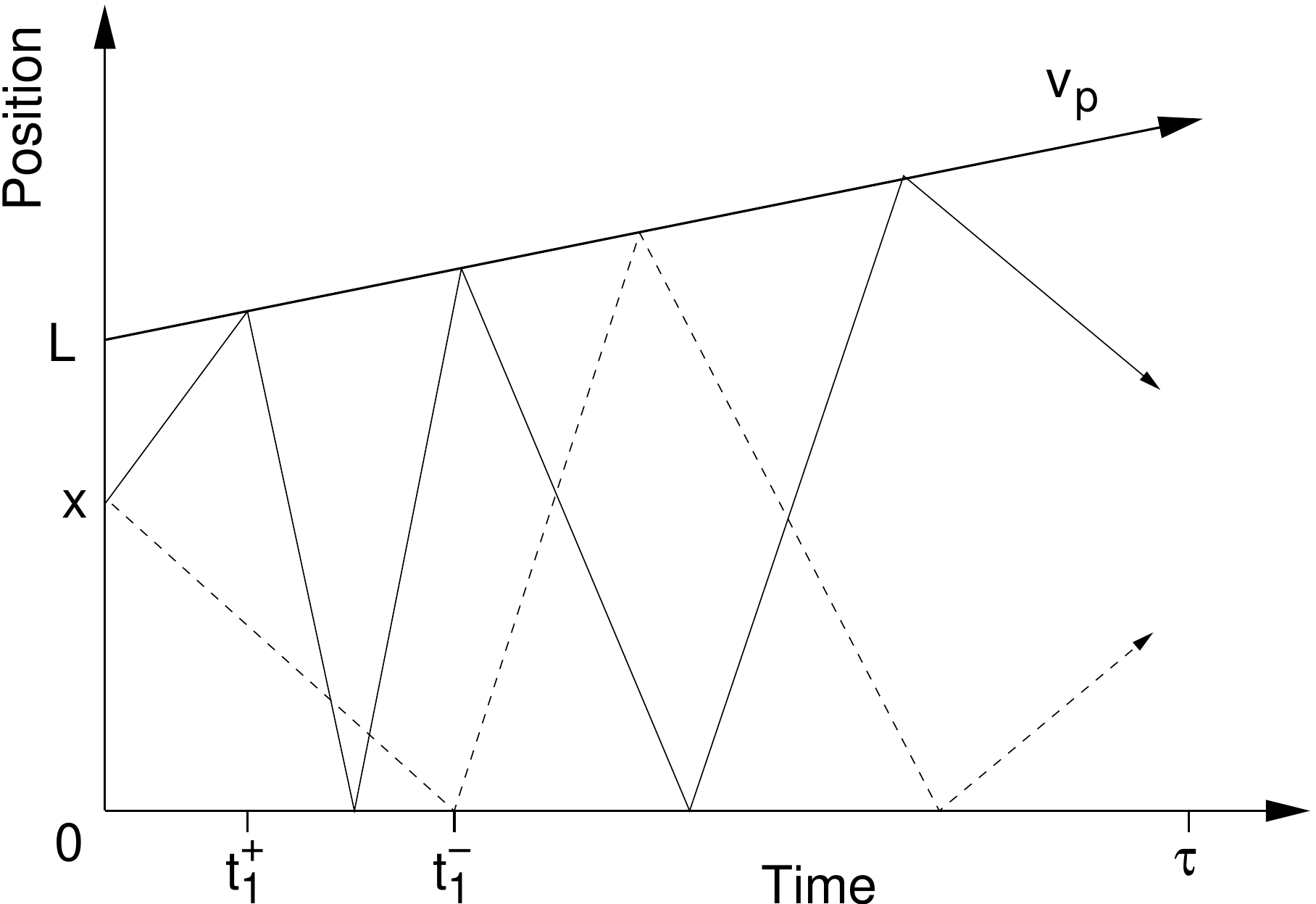}
\caption{\label{Diagramm}Position-time-diagram for the work process. The particle performs work by hitting the piston during the time of its movement $\tau$. Example trajectory for a positive (negative) initial velocity is denoted by a solid (dotted) line. Note the difference from the corresponding diagram in \cite{Lua}. Here the particle trajectories have varying slopes due to the isothermal boundaries.}
\end{center}
\end{figure}

The isothermal ideal gas is subjected to the following thermodynamic process (see Fig.~\ref{Diagramm}). In the initial state the gas is in equilibrium with the heat bath, confined to the (one-dimensional) `volume' $L$ by a fixed position of the piston. An individual gas particle samples its velocity from a Maxwell-Boltzmann distribution $\rho_{MB}(v)$ and its position from the uniform distribution $1/L$. Then the external work process begins by moving the piston outwards at constant speed $v_p$ for a time period $\tau$. Work is performed when gas particles collide with the piston during its movement. (The retreating piston then does a negative amount of work on the gas.) When the piston stops the gas thermalizes to the final equilibrium state at volume $L+v_p\tau$. 

With regard to the previous discussion we identify the work parameter $\lambda$ as the position of the piston and the work rate as dependent on $\dot{\lambda}\equiv v_p$. Since the speed of the heat-transfer mechanism  remains fixed in this model, the piston velocity determines the reversibility or irreversibility of the process. In the limit $v_p\rightarrow 0$ with $\tau\rightarrow\infty$, we perform a quasistatic reversible expansion of the gas. In the converse case of large piston speed and short $\tau$, the bulk of the gas remains in the initial part of the container after the piston has stopped, although the volume is extended. Subsequent equilibration is only completed sometime after the full volume has been explored by the gas.

\subsection{Microscopic reversibility \label{Micro}}

It has been pointed out in Sec.~\ref{CrooksDeriv} that the validity of the non-equilibrium work theorem for stochastic dynamics in \cite{Crooks,Crooks2} relies on the condition of microscopic reversibility. The isothermal ideal gas model offers the opportunity to check Crooks' assumption (Eq.~(\ref{micro_rev})) against a physically motivated, realistic case. In order to determine the transition rates of Eq.~(\ref{db}) we have to discretize the phase space of the system into intervals of size $\Delta x$, $\Delta v$. As the internal energy is only changed by inelastic collisions with the boundaries, the discussion can be reduced to a single collision event in the vicinity of the piston.

First we consider the equilibrium case, where the work parameter (i.e.~the position of the piston) is fixed at $L$. The particle is in state $a$ if it occupies the phase space element $[v_a,v_a+\Delta v]\times[L-\Delta x,L]$ and in state $b$ if it is found in $[v_b,v_b+\Delta v]\times[L-\Delta x,L]$. We make $\Delta v$ infinitesimally small. Let $a$ and $b$ refer to the same position interval (but different velocities) before and after the collision. Obviously, in order to make the transition $a\rightarrow b$ by a bounce against the piston, $\textbf{v}_a$ and $\textbf{v}_b$ have opposite directions.

A transition rate $\omega(a\rightarrow b)=p(b,\Delta t|a,0)/\Delta t$ is given in terms of a conditional probability $p(b,\Delta t|a,0)$ for the system to be found in state $b$ at time $\Delta t$, given that it was in state $a$ at initial time $t_0=0$. In turn this conditional probability is determined by the joint probability $p(b,\Delta t;a,0)$ of finding the particle in state $a$ at time $0$ \textit{and} in state $b$ some time $\Delta t$ later:
\begin{eqnarray}
\label{cond_prob_def}
p(b,\Delta t|a,0)=\frac{p(b,\Delta t;a,0)}{p(a,0)}.
\end{eqnarray} 
The phase space probability distribution of an ideal gas at equilibrium is $\rho_{eq}(x,v)=1/L\,\rho_{MB}(v)$, thus the probability $p(a,0)$ is simply given by $p(a,0)=\rho_{eq}(x,v_a)\Delta x \Delta v$. On the other hand, the calculation of the joint probability $p(b,\Delta t;a,0)$ involves the following considerations. Initially the particle occupies state $a$. Relative to the piston its position is $\bar{\textbf{x}}=\textbf{x}-\textbf{L}$ with modulus $\bar{x}\in[0,\Delta x]$ and its velocity is $\textbf{v}_a$. After the bounce the new particle velocity is sampled from $\rho_B$ (Eq.~(\ref{rhoB})). Given that $v_a$ and $v_b$ are fixed, the transition $a\rightarrow b$ only occurs if $\bar{x}$ fulfils two conditions. First, the collision time $t_c\equiv\bar{x}/v_a \leq \Delta t$, since the collision has to take place within $\Delta t$. Second, when the particle travels at $\textbf{v}_b$ after the collision, it is found within $[0,\Delta x]$ at time $\Delta t$ only if $\Delta x \geq (\Delta t-t_c)v_b$. Consequently the joint probability $p(b,\Delta t;a,0)$ is determined as:
\begin{eqnarray}
\label{p_av_1}
p(b,\Delta t;a,0)&=&\langle\Theta(\bar{x})\Theta(\Delta x-\bar{x})\Theta(\Delta t-t_c')\nonumber\\
&&\times\Theta(\Delta x-(\Delta t-t_c')v_b')\nonumber\\
&&\times\delta(v_a-v_a')\delta(v_b-v_b')\rangle.
\end{eqnarray}
The average is taken with respect to the stochastic variables $\bar{x}$, $v_a'$, $v_b'$, which are sampled from the distributions $\rho_{eq}(L-\bar{x},v_a')$ and $\rho_B(v_b')$:
\begin{eqnarray}
\label{p_av_2}
p(b,\Delta t;a,0)&=&\rho_B(v_b)\Delta v\int_{0}^{\Delta x} \d \bar{x}\;\Theta(\Delta t - \bar{x}/v_a)\nonumber\\
&&\times\Theta(\Delta x-(\Delta t-\bar{x}/v_a)v_b)\nonumber\\
&&\times\rho_{eq}(L-\bar{x},v_a)\Delta v.
\end{eqnarray}
Since the particle position is uniformly distributed, $\rho_{eq}(L-\bar{x},v_a)$ can be taken out of the average and finally cancels when the conditional probability Eq.~(\ref{cond_prob_def}) is considered. One is essentially left with an integral over a product of two theta functions, whose result is presented in appendix \ref{app_2}. We obtain for the conditional probability $p(b,\Delta t|a,0)$:
\begin{eqnarray}
\label{cond_prob}
p(b,\Delta t|a,0)=A(v_a,v_b)e^{-v_b^2/2},
\end{eqnarray}
where $A(v_a,v_b)$ is a function symmetric in $v_a$, $v_b$ ($k_BT$ has been set to unity). The calculation of the conditional probability $p(a,\Delta t|b,0)$ for the transition $b\rightarrow a$ under time reversal follows the same steps and yields Eq.~(\ref{cond_prob}) with $v_a$, $v_b$ exchanged. It then follows that the ratio of transition rates between states $a$ and $b$ is equal to the Boltzmann factor of their energy difference. We are therefore reassured that, in the absence of external work, the model respects the detailed balance condition, Eq.~(\ref{db}).

In the non-equilibrium case, the movement of the piston has to be taken into account. In the context of the discussion in Sec.~\ref{CrooksDeriv}, the derivation of Crooks' microscopic reversibility condition, Eq.~(\ref{micro_rev}), is tantamount to fixing the work parameter (the piston position) at a succession of different values $L(t)=\{L+v_pt_1,...,L+v_p\tau\}$, but disregarding the momentum exchange resulting from the movement. With that omission, the detailed balance condition follows trivially for each fixed piston position $L+v_pt_i$ since we can simply repeat the calculations above with the new position $L=L+v_pt_i$. Equation~(\ref{micro_rev}) would then follow.

However, the transition rates in the non-equilibrium case have to include the dynamic change of the work parameter. We will show that this leads to a violation of the balance condition used in the derivation of Eq.~(\ref{micro_rev}). Let us assume that the ideal gas is in a macroscopic non-equilibrium state due to the movement of the piston. The probability distribution of the particle is thus given by a function $\rho(x,v)$, whose precise form is unknown. We consider the transition $a\rightarrow b$ during time step $\Delta t$ due to a collision event in the vicinity of the piston at initial time $t_0$. The position of the piston is $L'=L+v_pt_0$ and we can define state $a$ similar to the static case above as the volume element in front of the piston at $t_0$. Likewise state $b$ is defined as the volume element in front of the piston at time $t_0+\Delta t$:
\begin{eqnarray}
a&\equiv&[v_a,v_a+\Delta v]\times[L'-\Delta x,L'], \nonumber\\
b&\equiv&[v_b,v_b+\Delta v]\times[L'+v_p\Delta t-\Delta x,L'+v_p\Delta t].
\end{eqnarray}
As before we are interested in the conditional probability $p_{v_p}(b,t_0+\Delta t|a,t_0)$. For this purpose we transform to coordinates in the frame of the piston (all quantities referring to this frame will be denoted by an overbar):
\begin{eqnarray}
\label{v_trafo}
\bar{\textbf{v}}_a&=&\textbf{v}_a-\textbf{v}_p, \nonumber \\
\bar{\textbf{v}}_b&=&\textbf{v}_b-\textbf{v}_p.
\end{eqnarray}
As well as $\textbf{x}'=\textbf{x}-(\textbf{L}+\textbf{v}_pt)$. If we consider states $a$ and $b$ in piston coordinates: $\bar{a}=[\bar{v}_a,\bar{v}_a+\Delta v]\times[0,\Delta x]$, $\bar{b}=[\bar{v}_b,\bar{v}_b+\Delta v]\times[0,\Delta x]$, we realize that they refer to the same situation as in the static case treated above. Consequently the joint probability $p_{v_p}(\bar{b},t_0+\Delta t;\bar{a},t_0)$ assumes the form of Eq.~{\ref{p_av_1}. The average is now taken with respect to $\bar{x}$, $\bar{v}'_a$, $\bar{v}'_b$, which are sampled from $\rho(L'-\bar{x},\bar{v}'_a+v_p)$ and $\rho_B(\bar{v}'_b)$ (the latter distribution remains unchanged because $\bar{v}'_b$ is the velocity obtained from the moving piston). This average can be calculated as before, assuming that the non-equilibrium distribution $\rho(L'-\bar{x},\bar{v}'_a+v_p)$ becomes uniform with respect to $\bar{x}$ within the interval $[0,\Delta x]$ for small $\Delta x$. As a consequence $\rho(L'-\bar{x},\bar{v}'_a+v_p)$ can again be taken out of the average and cancels with the identical term in $p(\bar{a},t_0)$. The conditional probability $p_{v_p}(\bar{b},t_0+\Delta t|\bar{a},t_0)$ finally assumes the same form as Eq.~(\ref{cond_prob}). For the reverse transition $b\rightarrow a$ under time reversal, the same considerations lead to the result as in the static case. We therefore find, that in the frame of the piston, the detailed balance condition is indeed fulfilled: 
\begin{eqnarray}
\label{db_pframe}
\frac{p_{v_p}(\bar{b},t_0+\Delta t|\bar{a},t_0)}{p_{v_p}(\bar{a},t_0|\bar{b},t_0+\Delta t)}&=&\frac{A(\bar{v}_a,\bar{v}_b)}{A(\bar{v}_b,\bar{v}_a)}e^{-\bar{v}^2_b/2+\bar{v}^2_a/2}\nonumber\\
&=&e^{-(\bar{E}_b-\bar{E}_a)}.
\end{eqnarray}
However, the situation is different in the laboratory frame. Using the transformation Eqs.~(\ref{v_trafo}) in order to transform Eq.~(\ref{db_pframe}) to the lab frame, we see that detailed balance is violated:
\begin{eqnarray}
\frac{p_{v_p}(b,t_0+\Delta t|a,t_0)}{p_{v_p}(a,t_0|b,t_0+\Delta t)} &=&\frac{e^{-(v_b+v_p)^2/2}}{e^{-(v_a-v_p)^2/2}}.
\end{eqnarray}
Consequently Eq.~(\ref{micro_rev}) does not hold either. As a result we observe that, in the laboratory frame, the isothermal ideal gas model with a moving piston violates Crooks' microscopic reversibility condition. 

\subsection{The exponential work average \label{Sec_exp_w}}

In order to verify the non-equilibrium work theorem for the ideal gas with stochastic boundary conditions, the main task is to evaluate the exponential work average of the isothermal expansion process. The free energy difference on the other hand can be calculated from simple thermodynamic considerations: 
\begin{eqnarray}
\label{free_energy}
\Delta F=\ln\frac{Z_A}{Z_B}=N\ln\left(\frac{L}{L+v_p\tau}\right) . 
\end{eqnarray}
Throughout the calculations we set the temperature parameter $\beta=1$ without loss of generality. In the case of a non-interacting gas the partition functions $Z_A$ (initial state) and $Z_B$ (final equilibrated state) both factorise and one can effectively reduce the calculations to a single particle $N=1$. This gas molecule performs work if it can hit the piston during its movement, i.e. during the time period $\tau$. Since it obtains a new randomly chosen velocity upon reaching either the wall or the piston, the time $t_k$ for $k$ bounces to occur depends on all realized velocities $v_1,...,v_k$ such that $t_k=t_k(\{v_1,...,v_k\},x;L,v_p)$. Throughout this article we shall use the convention that the variables $v_i$ always refer to the velocity of the particle in the reference frame of the particular wall, where this velocity has been obtained. Also all calculations are performed with the modulus of the velocities. Since the new velocity of the particle after collision with the piston is thus given in the frame of the piston, the following recursion relation holds, as can easily be verified by inspecting the position-time diagram (Fig.~\ref{Diagramm}):
\begin{eqnarray}
\label{t_k_recursion}
t_k=\frac{L}{v_k-v_p}+\frac{v_k}{v_k-v_p}t_{k-1} .
\end{eqnarray}

As a consequence we have to distinguish between a positive or negative initial velocity $v_1$ since, in each case, different bounces contribute to the work average. For a positive initial velocity the odd-numbered bounces yield the work contribution and the time for the first bounce is $t^+_1=(L-x)/(v_1-v_p)$; for a negative sign the even bounces contribute with $t^-_1=x/v_1$. Hence the exponential work average can be written as:
\begin{eqnarray}
\label{exp_w}
\left\langle e^{-W}\right\rangle=\left\langle e^{0}\right\rangle_+ +\left\langle e^{0}\right\rangle_- +\left\langle e^{-W}\right\rangle_+ +\left\langle e^{-W}\right\rangle_- ,
\end{eqnarray}
where $+/-$ refers to a positive or negative initial velocity and the average is to be taken over the uniformly distributed initial position $x$ and all velocities $v_1,...,v_k$. The zero-work contributions to the exponential work average are then determined according to:
\begin{eqnarray}
\label{zero_work}
\left\langle e^{0}\right\rangle_+ &=&\frac{1}{L}\int_0^L \d x\int_0^\infty \d v_1\:\rho_{MB}(v_1) \nonumber\\
&&\times\left[\Theta(t^+_1-\tau)\Theta(t^+_1)+\Theta(-t^+_1)\right], \nonumber\\
\left\langle e^{0}\right\rangle_- &=&\frac{1}{L}\int_0^L \d x\int_0^\infty \d v_1\:\rho_{MB}(v_1)\int_0^\infty \d v_2\;\rho_B(v_2)\nonumber\\
&&\times\left[\Theta(t^-_2-\tau)\Theta(t^-_2)+\Theta(-t^-_2)\right] .
\end{eqnarray}
Here $\Theta(x)$ denotes the Heaviside theta function.
Recall that a negative amount of work is performed on the gas by the piston, since we consider an expanding volume. The total work $W$ is given by summation of the contributions due to the individual bounces against the piston. The single bounce contributions,
\begin{eqnarray}
\label{single_work}
w_i=-\Delta \textbf{p}\cdot \textbf{v}_p=-(v_i+v_{i+1})v_p+v_p^2,
\end{eqnarray}
are determined by the momentum transfer $\Delta \textbf{p}=\textbf{v}_i-(\textbf{v}_{i+1}+\textbf{v}_p)$, where we have set the particle mass to unity without loss of generality. Here, $\textbf{v}_{i+1}$ is always the new velocity of the particle after collision with the piston and is therefore given in the frame of the piston. The shift $-\textbf{v}_p$ takes this into account when considering the momentum transfer in the laboratory frame, leading to the term $+v_p^2$ in the work contribution Eq.~(\ref{single_work}). The $w_i$ are statistically independent random variables.

The average of the exponential work can be expressed as a series in the number of bounces $n$ with the piston:
\begin{eqnarray}
\label{wav+}
\left\langle e^{-W}\right\rangle_+ &=&\frac{1}{L}\sum_{n=1}^{\infty}\int_0^L \d x\int_0^{\infty} \d v_1...\int_0^{\infty} \d v_{2n+1} \times\nonumber\\
&&P_+\left(\{v_1,...,v_{2n+1}\},2n-1;\tau\right)e^{-\sum_{i=1}^n w_{2i-1}} \nonumber\\
\end{eqnarray} 
We introduce the joint probability distribution function $P_+\left(\{v_1,...,v_{2n+1}\},2n-1;\tau\right)$ which determines the probability of a particular realization of velocities $\{v_1,...,v_{2n+1}\}$ and of exactly $n$ bounces with the piston resulting, within the time period $\tau$:
\begin{eqnarray}
&P_+&\left(\{v_1,...,v_{2n+1}\},2n-1;\tau\right)\nonumber\\
&=&\rho_{MB}(v_1)\rho_B(v_2)...\rho_B(v_{2n+1})[\Theta(\tau-t^+_{2n-1}) \times\nonumber\\
&&\Theta(t^+_1)...\Theta(t^+_{2n-1})
-\Theta(\tau-t^+_{2n+1})\Theta(t^+_1)...\Theta(t^+_{2n+1})] \nonumber\\
\end{eqnarray}

An analogous expression holds for $\left\langle e^{-W}\right\rangle_-$:
\begin{eqnarray}
\label{wav-}
\left\langle e^{-W}\right\rangle_- &=&\frac{1}{L}\sum_{n=1}^{\infty}\int_0^L \d x\int_0^{\infty} \d v_1...\int_0^{\infty} \d v_{2n+2} \times\nonumber\\
&&P_-\left(\{v_1,...,v_{2n+2}\},2n;\tau\right)e^{-\sum_{i=1}^n w_{2i}}
\end{eqnarray} 
with the joint probability distribution
\begin{eqnarray}
&P_-&\left(\{v_1,...,v_{2n+2}\},2n;\tau\right) \nonumber\\
&=&\rho_{MB}(v_1)\rho_B(v_2)...\rho_B(v_{2n+2})[\Theta(\tau-t^-_{2n})\times\nonumber\\
&&\Theta(t^-_1)...\Theta(t^-_{2n})-\Theta(\tau-t^-_{2n+2})\Theta(t^-_1)...\Theta(t^-_{2n+2})]\nonumber\\
\end{eqnarray}

The difficulties in evaluating the exponential work average originate primarily from the integration over the theta function $\Theta(\tau-t_k(\{v_1,...,v_k\},x))$ where $t_k$ is given by the recursion relation (\ref{t_k_recursion}). Therefore we resort to a numerical investigation of the average, Eq.~(\ref{exp_w}), in Sec.~\ref{numerics} and in Sec.~\ref{FastLimit} below to an analytical but approximate evaluation, which tends to the exact answer in a well controlled limit.

\subsection{The limit of a fast moving piston \label{FastLimit}}

The assumption that at most one bounce takes place between particle and piston yields an approximation that is analytically tractable and becomes exact in the limit of a fast moving piston while the volume extension is small compared with the original volume. In this case, $L \gg v_p\tau\gg\tau$ where velocities are measured in units of the thermal velocity since $\beta\equiv m\equiv 1$. In Ref.~\cite{Lua} the one-bounce approximation validated the non-equilibrium work theorem for the adiabatic ideal gas expansion by considering the $n=1$ approximation of the work distribution. Here we establish this result by calculating the one-bounce approximation of the exponential work average Eq.~(\ref{exp_w}). According to the Jarzynski relation this average should yield the exponential of the free energy difference Eq.~(\ref{free_energy}):
\begin{eqnarray}
\label{exp_free}
e^{-\Delta F}=1+\frac{v_p\tau}{L} .
\end{eqnarray}
The $n=1$ approximation of the exponential work average consists of four contributions:
\begin{eqnarray}
\label{n1_av}
\left\langle e^{-W}\right\rangle_{n=1}=\left\langle e^{0}\right\rangle_+ +\left\langle e^{0}\right\rangle_- +\left\langle e^{-w_1}\right\rangle_+ +\left\langle e^{-w_2}\right\rangle_- .
\end{eqnarray}
Here $\left\langle e^{0}\right\rangle_+$ and $\left\langle e^{0}\right\rangle_-$ are given as in (\ref{zero_work}). The two one-bounce contributions, from Eqs.~(\ref{wav+}) and (\ref{wav-}), are:
\begin{eqnarray}
\label{n1_work}
\left\langle e^{-w_1}\right\rangle_+ &=&\frac{1}{L}\int_0^L \d x\int_0^\infty \d v_1\:\rho_{MB}(v_1)\int_0^\infty \d v_2\:\rho_{B}(v_2)\nonumber\\
&&\times\Theta(\tau-t^+_1)\Theta(t^+_1)e^{-w_1}, \nonumber\\
\left\langle e^{-w_2}\right\rangle_- &=&\frac{1}{L}\int_0^L \d x\int_0^\infty \d v_1\:\rho_{MB}(v_1)\int_0^\infty \d v_2\:\rho_B(v_2)\nonumber\\
&&\times\int_0^\infty \d v_3\:\rho_B(v_3)\Theta(\tau-t^-_2)\Theta(t^-_2)e^{-w_2}
\end{eqnarray}
In the appendix we derive the following results in the limit $L \gg v_p\tau\gg\tau$:
\begin{eqnarray}
\label{n1limits}
\left\langle e^{0}\right\rangle_+ +\left\langle e^{0}\right\rangle_- &\rightarrow& 1 ,\nonumber\\
\left\langle e^{-w_2}\right\rangle_- &\rightarrow& 0 .
\end{eqnarray}
In this section we calculate the dominant contribution $\left\langle e^{-w_1}\right\rangle_+$. This average takes the explicit form:
\begin{eqnarray}
\left\langle e^{-w_1}\right\rangle_+&=&\frac{1}{\sqrt{2\pi}L}\int_{0}^\infty \d v_2\; v_2\;e^{-(v_2-v_p)^2/2}\int_0^L dx\nonumber\\
&&\times\int_{v_p}^\infty \d v_1\;e^{-(v_1-v_p)^2/2}\Theta\left(\tau-\frac{L-x}{v_1-v_p}\right). \nonumber\\
\end{eqnarray}
The integrations over the $v_1$ and $v_2$ variables are independent. For $v_p \gg 1$ the integral over $v_2$ yields $\sqrt{2\pi}v_p$. Making the substitution $v=v_1-v_p$ and changing the argument of the theta function, we can calculate the remaining integral in a straightforward way:
\begin{eqnarray}
\left\langle e^{-w_1}\right\rangle_+&=&\frac{v_p}{L}\int_0^\infty \d v\;e^{-v^2/2}\int_0^L \d x\;\Theta(x-(L-v\tau)) \nonumber\\
&=&\frac{v_p\tau}{L}\int_0^{L/\tau}\d v\;v\;e^{-v^2/2}+v_p\int_{L/\tau}^\infty \d v\;e^{-v^2/2}\nonumber\\
&\rightarrow& \frac{v_p\tau}{L} .
\end{eqnarray}
The last limit holds for $L \gg v_p\tau$ such that, in combination with Eq.~(\ref{n1limits}), we obtain the expected result Eq.~(\ref{exp_free}).
\begin{figure}
\begin{center}
\includegraphics[width=7cm]{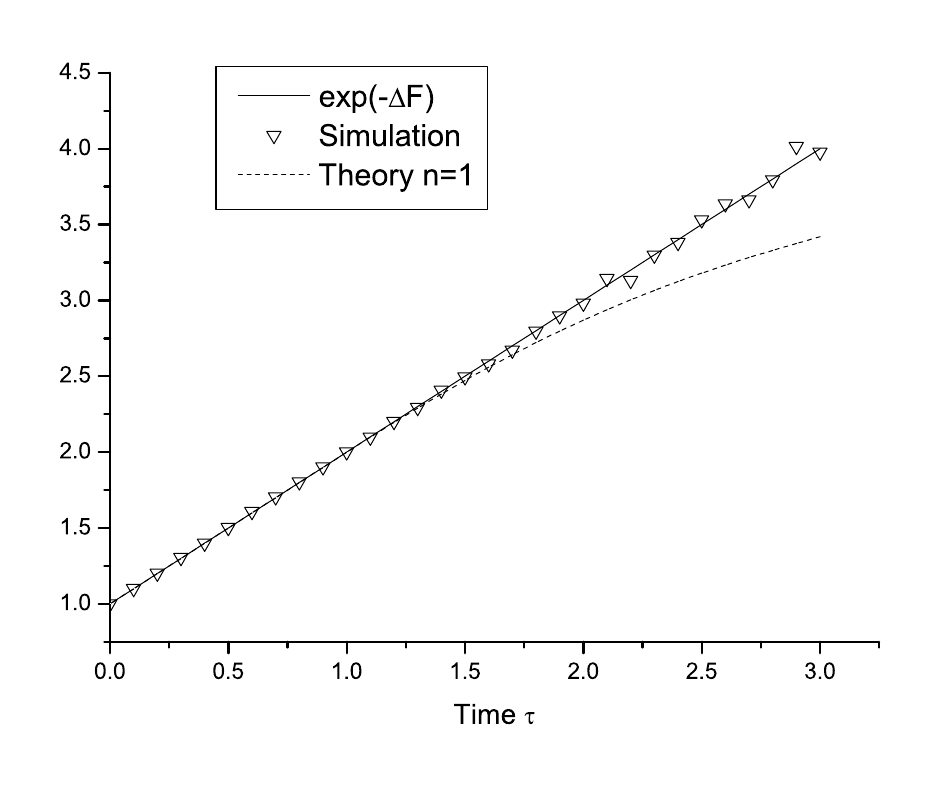}
\caption{\label{n1_fig} Comparison of the $n=1$ approximation Eq.~(\ref{n1_av}) (dotted line) with the exponential free energy difference $1+v_p\tau/L$ (solid line) for $v_p=1$. The triangles show the numerical results for the same parameter values. }
\end{center}
\end{figure}
Fig.~\ref{n1_fig} shows a plot of $\left\langle e^{-W}\right\rangle_{n=1}$ for $L=1$, $v_p=1$ together with Eq.~(\ref{exp_free}). According to our numerical results presented below a piston velocity in this regime drives the system sufficiently out of equilibrium. One notes that here the one-bounce approximation reproduces Jarzynski's relation for times up to $\tau\approx 1$, so that this approximation actually holds in a broader regime than the limits considered above. For larger times $\tau$ the single bounce is no longer sufficient to approximate the averaged exponential and higher order terms have to be taken into account.

\section{Numerical method \label{numerics}}

We have performed a numerical study of the isothermal ideal gas model in order to evaluate the exponential work average, Eq.~(\ref{exp_w}), for arbitrary numbers of bounces $n$. This thermostated `Molecular Dynamics' simulation essentially consists of picking random numbers from the velocity distributions $\rho_{MB}(v_1)$,  $\rho_{B}(v_i)$ and calculating the bouncing times $t_k$ (Eq.~(\ref{t_k_recursion})). When the velocity configuration allows the particle to hit the piston one or more times within $\tau$, the corresponding work values (Eq.~(\ref{single_work})) are recorded.
 
A general problem of the applicability of the non-equilibrium work theorem is the convergence of the average. Since the exponential $\exp[-\beta W]$ emphasizes small work values, one effectively has to sample the far left tail of the work distribution. In our model, the work performed on the system is always negative and the average is dominated by those events that lead to a large negative work contribution. A general discussion of the convergence problem can be found in \cite{Jarzynski3} with particular focus on the ideal gas and piston, but for an adiabatic work process (elastic collisions). It was shown that, in this particular case, the number of realizations needed in order to sample the dominant part of the average grows exponentially with the system size \cite{Lua} or, more generally, this number is proportional to the exponential of the averaged work that is dissipated during the reverse process \cite{Jarzynski3}. It is not obvious whether these results can be directly applied to the isothermal ideal gas model under consideration. In contrast to the adiabatic case, the occurrence of many collisions in a particular realization does not necessarily imply a large work contribution (a dominant event), as the work depends not only on the incoming velocity but also on the statistically independent outgoing velocity.
 
The numerical results below have been obtained with typically $10^6$ realizations per data point, which yields such excellent convergence of the exponential average that error bars have been omitted in the figures. All less relevant parameters are set to unity for simplicity and units are non-dimensional. This includes the length of the initial volume, $L=1$, and the inverse temperature, $\beta=1$, which sets the width of $\rho_{MB}$ and $\rho_B$. Accordingly the thermal velocity is unity as well so, for piston speeds $v_p>1$, we are in a regime where the tail of the initial velocity distribution contributes the work. The plots show the average exponential work for a given $v_p$ when $\tau$ is varied from zero up to the extended volume $v_p\tau=L$ (i.e.~the volume is doubled).

\subsection{Numerical results}

\begin{figure}
\begin{center}
\includegraphics[width=7cm]{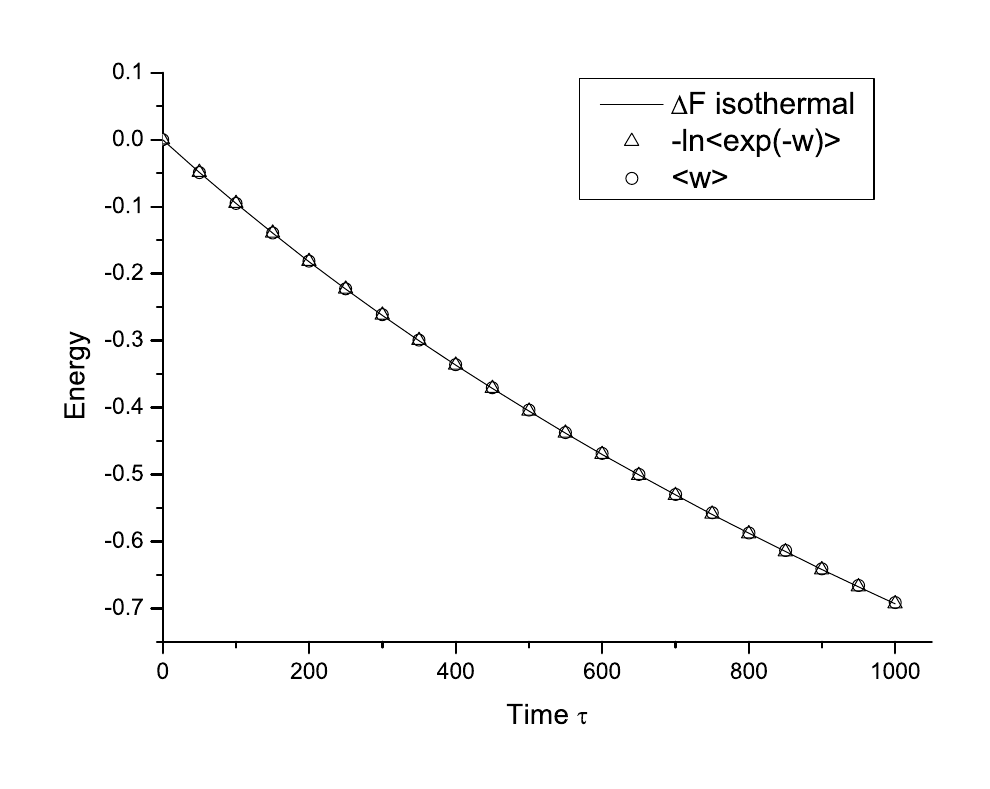}
\caption{\label{iso_rev_fig} Reversible isothermal expansion of the ideal gas for $v_p=0.001$. }
\end{center}
\end{figure}

\begin{figure}
\begin{center}
\includegraphics[width=7cm]{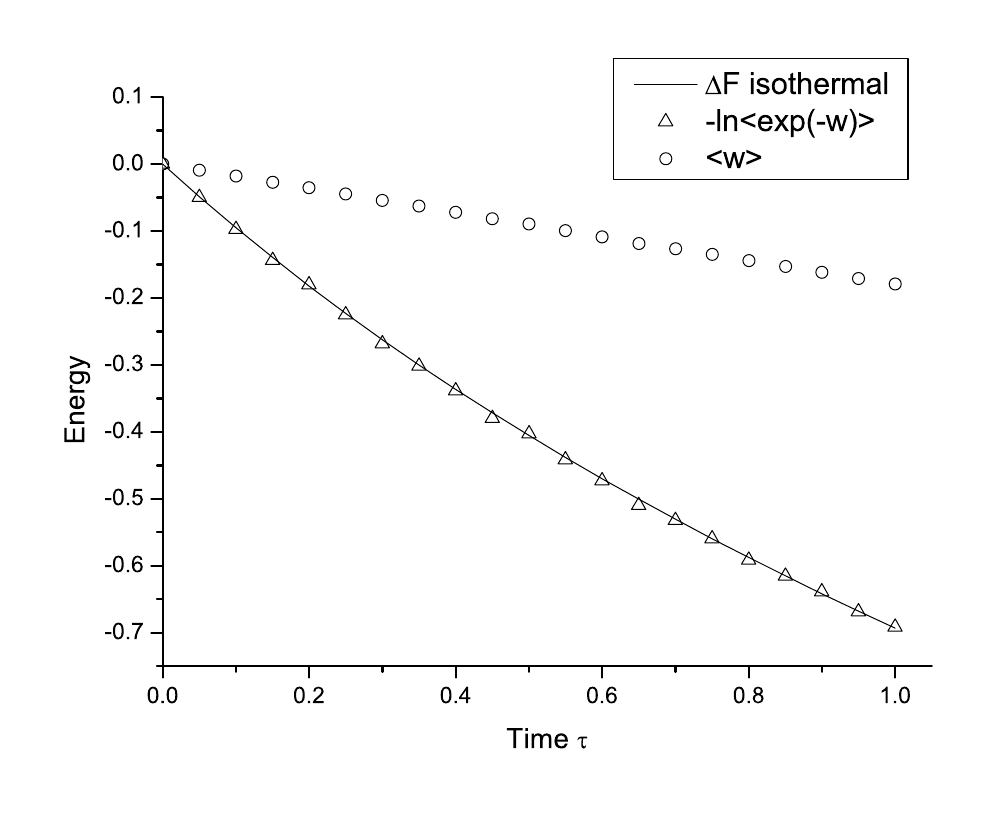}
\caption{\label{iso_irrev_fig} Irreversible isothermal expansion of the ideal gas for $v_p=1$. One notices the dissipated work as the difference between the average work and $\Delta F$. }
\end{center}
\end{figure}

\begin{figure}
\begin{center}
\includegraphics[width=7cm]{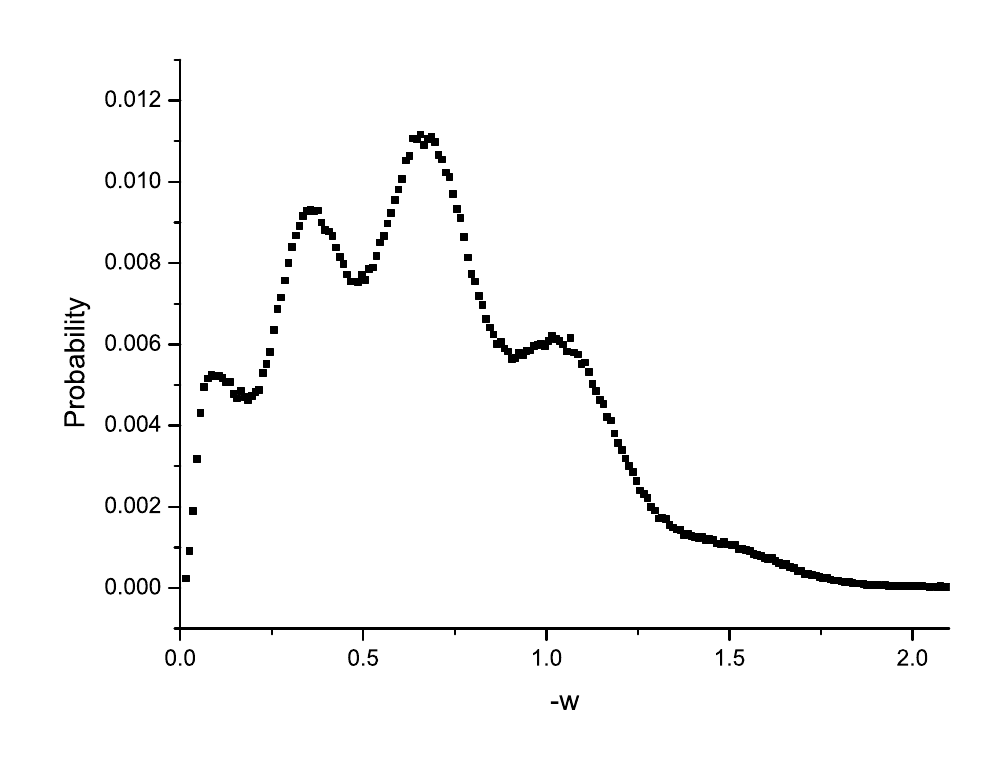}
\caption{\label{dist_fig} Work distribution of the isothermal expansion process for $v_p=0.1$.}
\end{center}
\end{figure}
We report the following results. In the limit of a very slow (quasistatic) expansion we obtain the isothermal free energy difference, Eq.~(\ref{free_energy}), from both the work average $\left<W\right>$, in accordance with the Second Law for a reversible process, and the exponential work average (see Fig.~\ref{iso_rev_fig}). The quasistatic regime is found at $v_p\leq0.001$. If we pull the piston at a higher speed the work average deviates noticeably from the free energy difference, indicating the onset of dissipation. The dissipated work is $W_d=\left< W\right>-\Delta F$ and we are effectively performing an irreversible non-equilibrium experiment. On the other hand the negative logarithm of the exponential work average still agrees with the isothermal free energy difference as predicted by the non-equilibrium work theorem (see Fig.~\ref{iso_irrev_fig}). This is the main result of our numerical investigation: that the Jarzynski relation holds in the non-equilibrium regime of our model ($v_p>0.001$) despite the fact that the model does not belong to the class of systems for which the theorem was derived, but has a more physically motivated coupling to its heat bath. The simulation shows excellent convergence up to $v_p\approx 1$. For higher piston speeds it becomes increasingly difficult to sample the tails of the velocity distributions. 

In Fig.~\ref{dist_fig}, we present the full distribution of work values determined numerically for $v_p=0.1$. The distribution exhibits a multi-peak structure, demonstrating that, in imposing only one constraint on the distribution, the non-equilibrium work theorem does not confine it to adopt a simple shape.

\subsection{The adiabatic piston model revisited}

An ideal gas with a piston was previously investigated by Lua and Grosberg \cite{Lua} for the case of adiabatic expansion, i.e.~perfectly elastic collisions at the boundaries. We can reproduce their adiabatic model by considering elastic, energy-conserving (and therefore deterministic) collisions instead of completely inelastic, stochastic ones. Consequently the probability distribution, Eq.~(\ref{rhoB}), is substituted by
\begin{eqnarray}
\label{rhoB_ad}
\rho_B(v_{out})=\delta(v_{out}-(v_{in}-v_p)) ,
\end{eqnarray}
when the incoming velocity is $v_{in}$. The shift $-v_p$ is due to our convention for the velocity variables $v_i$ (see Sec.~\ref{Sec_exp_w}) and is explained as follows. When the particle collides with the piston, $v_{in}$ is given in the lab frame, whereas $v_{out}$ refers to the velocity in the piston frame, therefore $v_{out}=v_{in}-v_p$. For the subsequent bounce against the fixed wall, $v_{in}$ is given in the piston frame, but $v_{out}$ refers to the velocity in the lab frame, hence again $v_{out}=v_{in}-v_p$ if the collision is elastic. Overall the distribution Eq.~(\ref{rhoB_ad}) is valid for bounces on the wall and on the piston side. However, when the initial velocity is negative, the first collision takes place with the fixed wall such that both incoming and out-going velocities refer to the lab frame. In this case only, $\rho_B(v_2)=\delta(v_2-v_1)$. Using these distributions, the bouncing times $t_k(\{v_1,...,v_k\},x)$ in Eq.~(\ref{t_k_recursion}) are reduced to the correct elastic counterpart $t_k(v_1,x)$ and the averages, Eqs.~(\ref{wav+}) and (\ref{wav-}), are evaluated by integrating over the initial velocity $v_1$ and initial position $x$. In this case the non-equilibrium work theorem has been proven to hold exactly for all parameter values \cite{Lua}.

We observe a particularly interesting feature of the non-equilibrium work theorem for the adiabatic expansion of the ideal gas, that is worth highlighting. In the quasistatic limit of a very slow moving piston, the work average $\left<W\right>$ yields the free energy difference of the adiabatic expansion of the gas as one would expect. The exponential work average, on the other hand, yields the free energy difference of the {\em isothermal} expansion of the gas. This can clearly be seen in Fig.~\ref{ad_rev_fig}.
\begin{figure}
\begin{center}
\includegraphics[width=7cm]{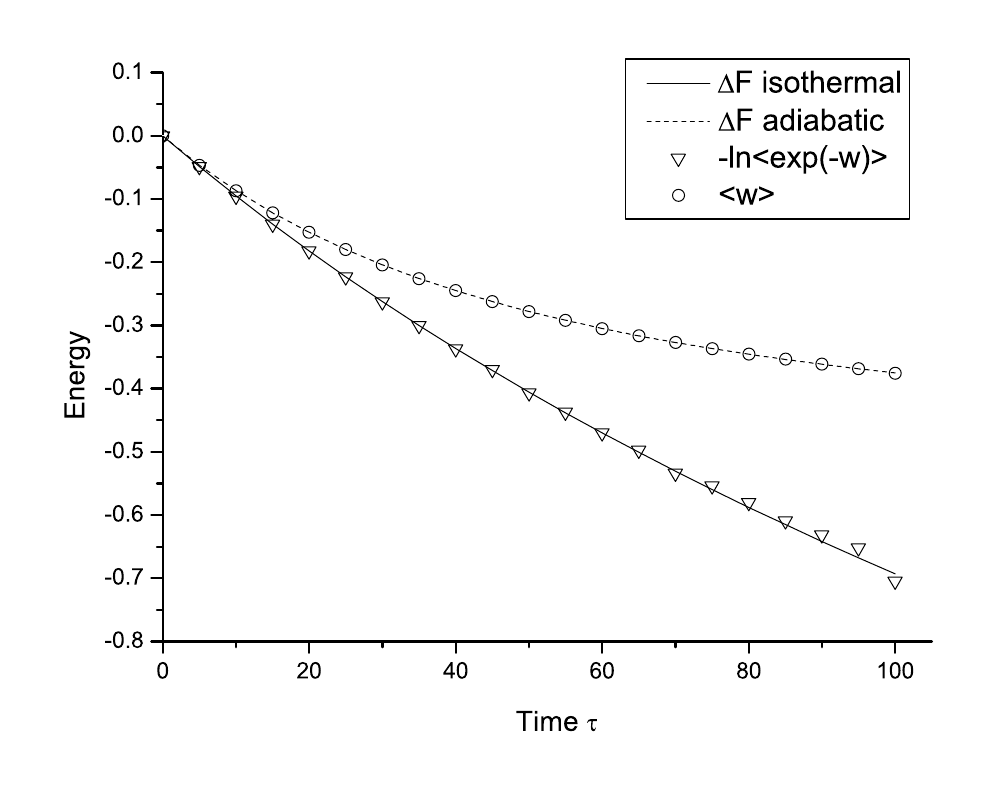}
\caption{\label{ad_rev_fig} Reversible adiabatic expansion of the ideal gas for $v_p=0.01$. Both the free energy difference for isothermal expansion and that for adiabatic expansion are obtained from the same work values.}
\end{center}
\end{figure}

\begin{figure}
\begin{center}
\includegraphics[width=7cm]{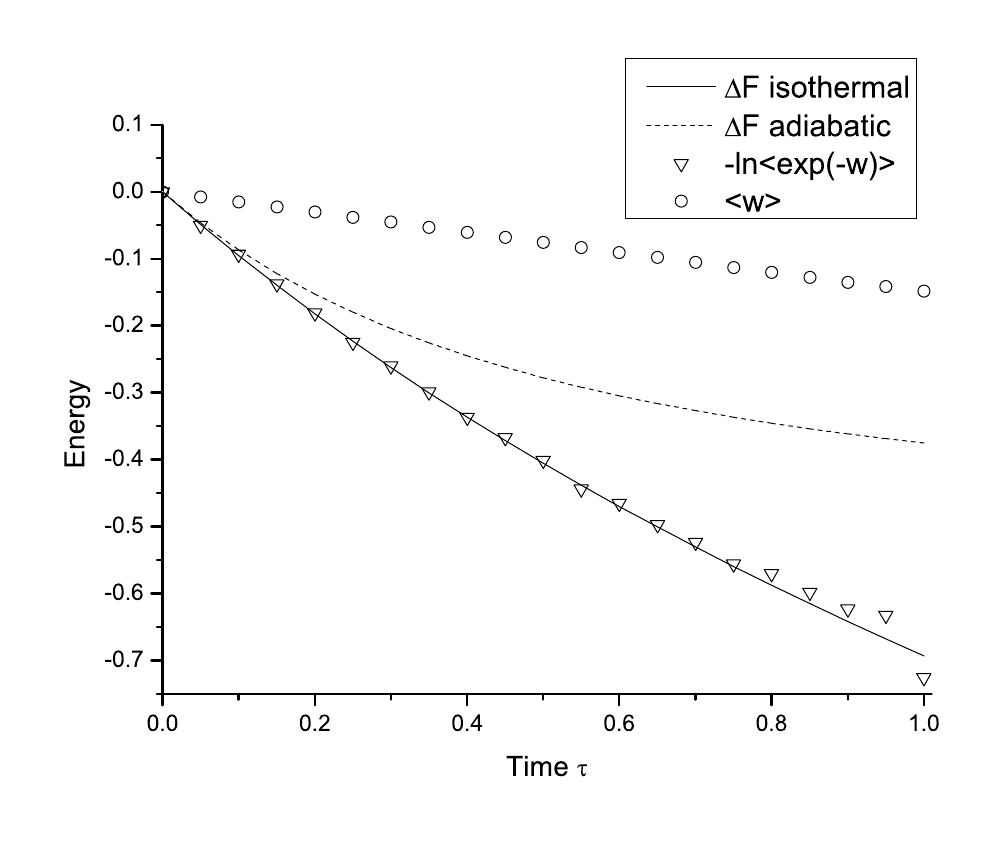}
\caption{\label{ad_irrev_fig} Irreversible adiabatic expansion of the ideal gas for $v_p=1$. The dissipated work is observed. }
\end{center}
\end{figure}
The averages $-\log\left<\exp(-W)\right>$ and $\left<W\right>$ respectively show excellent agreement with the isothermal free energy difference Eq.~(\ref{free_energy}) (solid line) and the adiabatic free energy difference (dotted line). The latter reads explicitly:
\begin{eqnarray}
\label{free_energy_ad}
\Delta F_{\rm ad}=\frac{1}{2}\left(\left(\frac{L}{L+v_p\tau}\right)^2-1\right).
\end{eqnarray}
Thus, by performing an adiabatic experiment, we obtain information about both an adiabatic and an isothermal system. If we consider the irreversible case for this adiabatic process, we observe again the dissipated work $W_d$ (see Fig.~\ref{ad_irrev_fig}) which is responsible for the deviation of $\left<W\right>$ from $\Delta F_{\rm ad}$. On the other hand $\Delta F$ (the isothermal result) can still be determined by evaluating the exponential work average. By application of the non-equilibrium work theorem we can perform an adiabatic or an isothermal experiment and obtain the same result, the isothermal free energy difference. From a numerical point of view the isothermal simulation proves slightly more advantageous because more sampling takes place during each realization of the protocol leading to a faster convergence of the exponential average.

\section{Conclusion}

The main result of our investigation is the validation, both numerically and analytically, of the non-equilibrium work theorem for the isothermal expansion of an ideal gas against a piston. Although the analytical calculation is restricted to the limit of a fast-moving piston and small volume extension, the simulation confirms the result for a wide range of parameter values. The two main characteristics of the model under consideration should be emphasized again. First, the isothermal model exhibits strong thermal coupling between system and heat reservoir, an important and more physically relevant extension to the ideal gas models previously discussed in the literature \cite{Lua,Bena}. Second, both the system and the heat reservoir depend on the work parameter $\lambda$, violating the assumptions in Jarzynski's original derivation \cite{Jarzynski2}. Furthermore it has been shown that microscopic reversibility is broken due to the moving and thermostatting piston, such that Crooks' derivation \cite{Crooks} (which assumes transition rates independent of $\dot{\lambda}$) does not hold either. We have thus identified a regime where Jarzynski's relation might have been expected to fail, however it appears that the validity of the non-equilibrium work theorem is not affected.

It is a pleasure to acknowledge helpful discussions with C. Jarzynski and comments by A. Adib. This work was funded by EPSRC Grant GR/T24593/01. RMLE is funded by the Royal Society.

\begin{appendix}

\section{The $n=1$ approximation}
We show the convergence of the averages $\left<e^0\right>_+ +\left<e^0\right>_- \rightarrow 1$ and $\left<e^{-w_2}\right>_- \rightarrow 0$ in the limit of a fast moving piston and small extended volume. In order to calculate the averages, we resolve the theta functions with respect to $v_1$ and obtain multiple dependent integrals.
 
First the zero-work contribution for positive initial velocity: \begin{eqnarray}
\label{app_zero+}
\left<e^0\right>_+=\frac{1}{\sqrt{2\pi}L}\int_0^L \d x\int_0^{(L-x)/\tau+v_p}\d v_1\;e^{-v_1^2/2} .
\end{eqnarray}
The zero-work contribution for negative initial velocity reads
\begin{eqnarray}
\label{app_zero-}
\left<e^0\right>_-&=&\frac{1}{\sqrt{2\pi}L}\int_0^L \d x\int_{L/\tau+v_p}^\infty \d v_2\;v_2\;e^{-v_2^2/2}\nonumber\\
&&\times\int_0^{(v_2x/(v_2-v_p)\tau-L)}\d v_1\;e^{-v_1^2/2}\nonumber\\ &&+\frac{1}{2}\int_0^{L/\tau+v_p}\d v_2\;v_2\;e^{-v_2^2/2},
\end{eqnarray}
and the one-bounce contribution for negative initial velocity is given as
\begin{eqnarray}
\label{app_avw-}
\left<e^{-w_2}\right>_-&=&\frac{1}{\sqrt{2\pi}L}\int_0^L \d x\int_{L/\tau+v_p}^\infty \d v_2\;v_2\;e^{-(v_2-v_p)^2/2}\nonumber\\
&&\times\int_{v_2x/((v_2-v_p)\tau-L)}^\infty \d v_1\;e^{-v_1^2/2}\nonumber\\
&&\times\int_0^\infty \d v_3\;v_3\;e^{-(v_3-v_p)^2/2}.
\end{eqnarray}
We discuss the limiting behaviour as follows. The integration over $v_1$ always leads to the error function (Eq.~(\ref{app_zero+}) and Eq.~(\ref{app_zero-})) or to a sum of a constant and the error function (Eq.~(\ref{app_avw-})). These expressions are always bounded by a constant for arbitrary values of the argument. For Eq.~(\ref{app_zero+}) we immediately obtain, in the limit of small $\tau$ and large $v_p$: $\left<e^0\right>_+ \rightarrow 1/2$.
 
For the two other cases we see that the integrand of the second integration is multiplied by a bounded term. Thus the convergence of both averages depends only on the $v_2$-integration. If we note that $x\exp(-x^2)$ is already small for $x>1$, we see that the first term in Eq.~(\ref{app_zero-}) is clearly vanishing in the considered limit and the result is $\left<e^0\right>_- \rightarrow 1/2$. In the case of Eq.~(\ref{app_avw-}) we simply observe that there is an additional shift $v_2-v_p$ in the argument of the exponential such that this integrand decays to zero even faster than the first term in Eq.~(\ref{app_zero-}). As a result the relations Eq.~(\ref{n1limits}) are valid and the single bounce approximation analytically confirms the non-equilibrium work-theorem in the regime $L \gg v_p\tau\gg\tau$.

\section{The conditional probability $p(b,\Delta t|a,0)$ \label{app_2}}

The integration over the product of the two theta functions in Eq.~(\ref{p_av_2}) yields the result:
\begin{eqnarray}
\label{int_result}
&&\int_{0}^{\Delta x}\d \bar{x}\;\Theta(\Delta t - \bar{x}/v_a)\Theta(\Delta x-(\Delta t-\bar{x}/v_a)v_b)\nonumber\\
&=&\Theta(v_a-\Delta x/\Delta t)\Theta(v_b-\Delta x/\Delta t)\nonumber\\
&&\times\Theta(\Delta x/v_a+\Delta x/v_b-\Delta t)[\Delta x-(\Delta t-\Delta x/v_b)v_a]\nonumber\\
&&+\Theta(v_a-\Delta x/\Delta t)\Theta(\Delta x/\Delta t-v_b)\Delta x\nonumber\\
&&+\Theta(\Delta x/\Delta t-v_a)\Theta(v_b-\Delta x/\Delta t)\Delta x \,v_a/v_b\nonumber\\
&&+\Theta(\Delta x/\Delta t-v_a)\Theta(\Delta x/\Delta t-v_b)\Delta t \,v_a\nonumber\\
&\equiv&B(v_a,v_b;\Delta x, \Delta t).
\end{eqnarray}
The joint probability $p(b,\Delta t;a,0)$ then reads:
\begin{eqnarray}
p(b,\Delta t;a,0)&=&\rho_{eq}(x,v_a)\rho_B(v_b)\Delta v^2 B(v_a,v_b;\Delta x,\Delta t).\nonumber\\
\end{eqnarray}
Dividing by $p(a,0)=\rho_{eq}(x,v_a)\Delta x\Delta v$ we thus obtain the conditional probability $p(b,\Delta t|a,0)$:
\begin{eqnarray}
p(b,\Delta t|a,0)=\rho_B(v_b)\frac{\Delta v}{\Delta x} B(v_a,v_b;\Delta x,\Delta t).
\end{eqnarray}
Finally this can be rewritten in the form of Eq.~(\ref{cond_prob}) if we consider that $\rho_B(v_b)=v_b\exp[-v_b^2/2]$ (for $k_BT$ set to unity) and
\begin{eqnarray}
A(v_a,v_b)\equiv\frac{\Delta v}{\Delta x}v_b B(v_a,v_b;\Delta x,\Delta t).
\end{eqnarray}
From Eq.~(\ref{int_result}) we see that $v_b B(v_a,v_b;\Delta x,\Delta t)$ is invariant under the exchange $v_a\leftrightarrow v_b$.

\end{appendix}


\begin{thebibliography}{}

\bibitem{Ritort} F. Ritort, \textit{Poincar\'e Seminar} 2, 195 (2003)
\bibitem{Park} S. Park and K. Schulten, \textit{J. Chem. Phys.} 120, 5946 (2004)
\bibitem{Jarzynski} C. Jarzynski, \textit{Phys. Rev. Lett.} 78, 2690 (1997)
\bibitem{Liphardt} J. Liphardt et al., \textit{Science} 296, 1832 (2002)
\bibitem{Cohen} E.G.D. Cohen and D. Mauzerall, \textit{J. Stat. Mech.}, P07006 (2004)
\bibitem{Lua} R.C. Lua and A.Y. Grosberg, \textit{J. Phys. Chem. B} 109, 6805 (2005)
\bibitem{Bena} I. Bena, C. Van den Broeck and R. Kawai, \textit{Europhys. Lett.} 71, 879 (2005)
\bibitem{Jarzynski2} C. Jarzynski, \textit{J. Stat. Mech.}, P09005 (2004)
\bibitem{Crooks} G.E. Crooks, \textit{J. Stat. Phys.} 90, 1481 (1998)
\bibitem{Crooks2} G.E. Crooks, \textit{Phys. Rev. E} 60, 2721 (1999)
\bibitem{Kampen} N.G. Van Kampen, \textit{Stochastic Processes in Physics and Chemistry} (North-Holland, Amsterdam, 1985)
\bibitem{Klein} M.J. Klein, \textit{Phys. Rev.} 97, 1446 (1955)
\bibitem{Evans} R.M.L. Evans, \textit{Phys. Rev. Lett.} 92, 150601 (2004)
\bibitem{Jarzynski3} C. Jarzynski, cond-mat/0603185
\bibitem{Silbey} S. Press\'e and R. Silbey, \textit{J. Chem. Phys.} 124, 054117 (2006)
\bibitem{Lebowitz} J.L. Lebowitz and H. Spohn, \textit{J. Stat. Phys.} 19, 633 (1978)

\end{thebibliography}
\end{document}